\RequirePackage{lineno} 
\documentclass[aps,prc,superscriptaddress,nofootinbib,showpacs,floatfix,twocolumn]{revtex4}
\usepackage{graphicx} 
\usepackage{float} 
\usepackage{amssymb}
\usepackage{url}
\usepackage{cancel}
\usepackage{MnSymbol}
\usepackage{multirow}
\usepackage{bm}
\usepackage{color}
\definecolor{dgreen}{cmyk}{1.,0.,1.,0.1}        % dark green
\definecolor{orange}{cmyk}{0.,0.353,1.,0.}    % orange

\usepackage{color}
\usepackage{harpoon}
\definecolor{my}{rgb}{1, 0, 0}
\begin{document} % do not change

\title{Generic algorithm for multi-particle cumulants of azimuthal correlations in high energy nucleus collisions}
\author{Zuzana Moravcova} 
\affiliation{Niels Bohr Institute, Blegdamsvej 17, 2100 Copenhagen, Denmark}
\author{Kristjan Gulbrandsen}
\email{gulbrand@nbi.dk}
\affiliation{Niels Bohr Institute, Blegdamsvej 17, 2100 Copenhagen, Denmark}
\author{You Zhou}
\email{you.zhou@cern.ch}
\affiliation{Niels Bohr Institute, Blegdamsvej 17, 2100 Copenhagen, Denmark}

\date{\today}

\begin{abstract}

Multi-particle cumulants of azimuthal angle correlations have been compelling tools to probe the properties of the Quark-Gluon Plasma (QGP) created in the ultra-relativistic heavy-ion collisions and the search for the QGP in small collision systems at RHIC and the LHC. However, only very few of them are available and have been studied in theoretical calculations and experimental measurements, while the rest are generally very interesting, but their direct implementation was not feasible. In this paper, we present a generic recursive algorithm for multi-particle cumulants, which enables the calculation of arbitrary order single and mixed harmonic multi-particle cumulants. Among them, the new 10-, 12-, 14-, and 16-particle cumulants of a single harmonic, named $c_{n}\{10\}$, $c_{n}\{12\}$, $c_{n}\{14\}$, and $c_{n}\{16\}$, and the corresponding $v_n$ coefficients, will be discussed for the first time. Our Monte Carlos studies show that these new multi-particle cumulants can be readily used along with updates to the generic framework of multi-particle correlations to a very high order. Finally, we propose a particular series of mixed harmonic multi-particle cumulants, which measures the general correlations between any moments of different flow coefficients. The predictions of these new observables are shown based on an initial state model {\tt MC-Glauber}, a toy Monte Carlo model, and the {\tt HIJING} transport model for future comparisons between experimental data and theoretical model calculations. The study of these new multi-particle cumulants in heavy-ion collisions will significantly improve the understanding of the joint probability density function which involves both different harmonics of flow and also the symmetry planes. This will pave the way for more stringent constraints on the initial state and help to extract more precisely information on how the created hot and dense matter evolves. Meanwhile the efforts applied to small systems could be very helpful in the understanding of the origin of the observed collectivity at RHIC and the LHC.

\end{abstract}
\maketitle

%\linenumbers

\section{Introduction}\label{sec:1}

The properties of an extreme state of matter, the quark-gluon plasma (QGP), are studied by colliding heavy ions at BNL's Relativistic Heavy Ion Collider (RHIC) and at CERN's Large Hadron Collider (LHC), allowing us to recreate this matter in the laboratory.
One of the popular approaches in the exploration of QGP properties is to study the anisotropic flow phenomena~\cite{Ollitrault:1992bk}. Anisotropic flow is quantified by flow coefficients $v_n$ along with their corresponding flow symmetry planes $\Psi_n$ in the Fourier series decomposition of the azimuthal particle distribution~\cite{Voloshin:1994mz}:
\begin{equation}
\frac{dN}{d\varphi} \propto 1 + 2\sum_{n=1}^{\infty} v_{n} \cos [n(\varphi-\psi_{n})],
\label{eq:Fourier}
\end{equation}
where $\varphi$ is the azimuthal angle of emitted particles. The flow coefficient $v_{n}$ and its corresponding flow symmetry plane $\Psi_n$ defines the $n$-th order flow-vector as ${\bf V}_{n} = v_{n}\,e^{in\Psi_{n}}$. Alternatively, the expression of anisotropy can be generally given by the joint probability density function in terms of the various harmonics of flow coefficients and their corresponding flow symmetry planes as:
\begin{eqnarray}
&& P(v_{m}, v_{n}, ..., \Psi_{m}, \Psi_{n}, ...)  \nonumber\\
&& ~~ =  \frac{1}{N_{event}} \frac{dN_{event}} {{\rm d} v_{m} \,{\rm d} v_{n} \cdot \cdot \cdot {\rm d}\Psi_{m} \, {\rm d}\Psi_{n} \cdot \cdot ~\cdot }
\label{eq:pdf}
\end{eqnarray}

The investigations on the joint $p.d.f.$ started from the study of multi-particle cumulants of single harmonic. Based on the experimental measurements of these multi-particle cumulants~\cite{Acharya:2018lmh}, the probability density function ($p.d.f.$) of a single harmonic $v_n$, called $P(v_2)$, can be constructed. The results are in a good agreement with the one obtained via the Bayesian unfolding approach~\cite{Sirunyan:2017fts, Aad:2013xma}. Both of the results show that $P(v_2)$ is well described by the Elliptic-Power function~\cite{Yan:2014afa} instead of Bessel-Gaussian function~\cite{Voloshin:2007pc}, which has been widely used before as the potential description of the underlying $p.d.f.$ of $v_2$.
Extensive experimental measurements of the transverse momentum and pseudorapidity dependence of anisotropic flow with charged and identified hadrons yield strong constraints on the theoretical models. The successful description of experimental flow measurements by hydrodynamic calculations have led to a broad and in-depth knowledge of the initial conditions and properties of the created hot/dense QCD matter. It suggested that the created QGP matter at ultra-relativistic heavy-ion collisions behaves as nearly perfect fluid~\cite{Heinz:2013th, Luzum:2013yya, Huovinen:2013wma, Shuryak:2014zxa, Song:2013gia, Dusling:2015gta, Song:2017wtw} with an extremely small specific shear viscosity $\eta/s$ close to the quantum limit of $1/(4\pi)$~\cite{Kovtun:2004de}.
However, to further pin down the uncertainty of the extracted $\eta/s$, especially its temperature dependence during the system expansion, more precise study on the initial conditions is highly necessary, but can not be obtained from the measurement of anisotropic flow coefficients alone.

While most of the studies mentioned above concentrate on the fluctuations of single harmonic flow and the corresponding symmetry planes, further study of correlations between different order flow-vectors sheds new light on the joint $P(v_{m}, v_{n}, ..., \Psi_{m}, \Psi_{n}, ...)$ function, which eventually will help to understand both the initial conditions and dynamic evolution of the QGP. The correlation between different order symmetry planes was investigated with multi-particle correlations~\cite{ALICE:2011ab, Acharya:2017zfg, Sirunyan:2019izh,Acharya:2019uia,Acharya:2020taj}. Meanwhile, symmetric cumulants, $SC(m,n)$~\cite{Bilandzic:2013kga}, made it possible for the first time to quantify the correlations between the second-order moments of flow coefficients, $v_{m}$ and $v_{n}$, experimentally. It has been found that $SC(m,n)$ has a unique sensitivity to the initial conditions of the QGP and can distinguish between various scenarios of the temperature dependence of $\eta/s$ in hydrodynamic and transport models~\cite{Bilandzic:2013kga, Bhalerao:2014xra, Niemi:2015qia, Aad:2015lwa, Zhu:2016puf}. 
It is an open question at the moment how the joint underlying $p.d.f.$ including different order symmetry planes and harmonics are described% and, additionally, if these correlations between different order symmetry planes and harmonics modify the single harmonic $p.d.f.$'s. 
Nevertheless, the study of not only multi-particle cumulants of a single harmonic but also the correlations between different order flow vectors will begin to answer this question. More investigations concerning correlations involving more than two different flow harmonics or higher order moments of flow coefficients along with their symmetry planes can further improve our understanding of the joint $p.d.f.$ and thus lead to new insights into the nature of the fluctuations of the created QGP in heavy-ion collisions.

%In addition, it was realized that the produced particles at different transverse momentums, $p_{\rm T}$, and pseudorapidity, $\eta$, might not share a common flow angle or symmetry plane. Such transverse momentum and pseudorapidity dependent flow vector fluctuate event-by-event, which also breaks the factorizations of the flow harmonics. It was examined via constructed 2- and multi-particle correlations of the same order of flow coefficient, e.g. $v_{n}[2]/v_{n}\{2\}$~\cite{} and factorization ratio $r_{n}$~\cite{ALICE-JHEP, CMS:2013bza, Khachatryan:2015oea}. These measurements were nicely predicted or reproduced by hydrodynamic calculations, and are found to be sensitive to either the initial-state density fluctuations and/or the shear viscosity of the expanding fireball medium~\cite{Heinz:2013bua, Gardim:2012im, Kozlov:2014hya}. 

On the other hand, the study of flow phenomena in small collision systems at  RHIC and the LHC is particularly interesting. Among many important measurements, the observation of anisotropic flow in high multiplicity events of small collision systems has attracted a lot of attention while the underlying mechanism is still under intense debate. The measurements of multi-particle cumulants of a single harmonic as well as mixed harmonics, e.g. symmetric and asymmetric cumulants, have been found to be extremely useful in determining whether the observed flow in small systems is attributed to final state interactions, such as hydrodynamic expansion~\cite{Bozek:2011if,Qin:2013bha,Weller:2017tsr,Schenke:2014zha}, parton cascades~\cite{Bzdak:2014dia,Ma:2014pva}, hadronic rescattering~\cite{Zhou:2015iba}, or a rope and shoving mechanism~\cite{Bierlich:2017vhg}, or it originates from initial state effects related to gluon saturation~\cite{Dusling:2012iga,Dusling:2012cg,Dusling:2013qoz,Dusling:2014oha,Dumitru:2014dra,Dumitru:2014vka,Noronha:2014vva,Schenke:2016lrs}, or it can be explained by the combinations of both initial and final state effects~\cite{Mantysaari:2017cni}. However, the critical challenge for these measurements is to remove the non-flow contamination which originates from the azimuthal correlations not associated with the common flow symmetry planes but rather from jets and resonance decays, etc. It is almost negligible for non-peripheral heavy-ion collisions but has a large influence on flow measurements in small collision systems. The development of the sub-event method of multi-particle cumulants~\cite{Jia:2017hbm,Huo:2017nms} has made progress towards this challenge, but is still far from sufficient. Another natural choice is using multi-particle cumulants of a higher order which further suppresses the non-flow contamination in multi-particle correlations. At the same time, they further increase the sensitivity to the event-by-event fluctuating initial conditions which has not been studied in detail yet. This is partly due to the implementation of higher order cumulants not being readily available. 

In this paper, we will take a step further in this direction with a newly established generic algorithm for arbitrarily high order multi-particle cumulants. More specifically, we, for the first time, propose and study the 10-, 12-, 14-, and 16-particle cumulants of a single harmonic, called $c_{n}\{10\}$, $c_{n}\{12\}$, $c_{n}\{14\}$, and $c_{n}\{16\}$, and their corresponding $v_n$ coefficients, $v_{n}\{10\}$, $v_{n}\{12\}$, $v_{n}\{14\}$, and $v_{n}\{16\}$. Furthermore, this generic algorithm is valid for any combination of mixed harmonic multi-particle cumulants. Based on this algorithm, a new series of flow observables are discussed using Monte Carlo model simulations. Together with the study of higher order cumulants, the study of higher order single and mixed harmonic cumulants could contribute significantly to the efforts of constraining the joint $p.d.f.$. The application to small collision systems provides a new window to probe the origins of collectivity in the near future.

%$observable of Flow Moment Correlation $MHC$ which enables the possibility to measure correlations between any moments of $v_n$ and $v_m$, and allows to study the correlations involving more than two different flow coefficients. Using various initial state models and different settings of $\eta/s$, we will illustrate that $MHC$ has the potential to open a new window into the nature of fluxing initial conditions and the dynamic evolution of the created QGP in heavy-ion collisions.

This paper is divided into specific sections. Section~\ref{sec:model} introduces the {\tt MC-Glauber}, toy Monte Carlo ({\tt toyMC}) and {\tt HIJING} transport models used to investigate these high order single and mixed harmonic cumulants. The new proposed algorithm of general multi-particle cumulants is introduced in section~\ref{sec:ga}. Examples of higher order cumulants of a single harmonic and mixed harmonics are discussed in sections~\ref{sec:single} and \ref{sec:mixed}. Finally, section~\ref{sec:summary} provides a summary of the results.

\section{The model and the setup of the calculations}
\label{sec:model}

In this section we introduce the Monte Carlo models used to study the multi-particle cumulants. Firstly, a toy Monte Carlo ({\tt toyMC}) is used to validate our generic algorithm and the general formulas (i.e. involving any set of harmonics) of various multi-particle cumulants. While the general joint $p.d.f.$, $P(v_{m}, v_{n}, ..., \Psi_{m}, \Psi_{n}, ...)$ can contain any number of correlations, for the purpose of validation we use a simplified static $p.d.f.$ of a few harmonics such that the angular distribution of each particle, $f(\varphi)$, is identical. $f(\varphi)$ is shown in Eq.~(\ref{eq:pdf}) where it is parametrized with only dominant contributions from the second and third harmonics, $v_2$ and $v_3$, without contributions from their corresponding flow symmetry planes, $\Psi_{2}$ and $\Psi_3$. We set the input value of $v_{2} = 0.10$ and $v_3 = 0.05$ for all events.
\begin{equation}
f(\varphi) = \frac{1}{2\pi}\bigg[1+ 2 \,v_2 \, \cos(2\varphi) + 2 \,v_3 \, \cos(3\varphi) \bigg]
\label{eq:pdfv2}
\end{equation}

In addition, in order to probe the initial conditions in heavy-ion collisions, the Monte Carlo Glauber model~\cite{Miller:2007ri, Loizides:2014vua} is used. This model is commonly used for calculations of geometric quantities in the initial state of heavy-ion collisions, such as the impact parameter, $b$, and the initial anisotropy coefficient, $\varepsilon_{n}= \frac{\sqrt{\langle r^{n} \sin (n \phi) \rangle ^{2} 
+ \langle r^{n} \cos (n \phi) \rangle ^{2} }} {\langle r^{n} \rangle }$. 
The version we use is based on the implementation of the PHOBOS Glauber Monte Carlo~\cite{Miller:2007ri, Loizides:2014vua} and has been used in several previous studies~\cite{ALICE:2011ab, ALICE:2016kpq}.
With the simple assumption that $v_n = \kappa_n \, \varepsilon_n$ (for $n=2, 3$), where $\kappa_n$ is a scaling number that depends on the properties of the produced matter, we can perform the calculations of multi-particle cumulants using $\varepsilon_n$ and then scale the results to obtain the multi-particle cumulants that involve $v_n$ coefficients. Except for the normalized flow observable, which we will introduce in the next section, the $\kappa$ parameter cancels out. Thus, we can give a direct prediction for the future experimental measurements based on initial state calculations using the {\tt MC-Glauber} model.

For more careful physics discussions on the newly studied multi-particle mixed-harmonic cumulants, the Heavy Ion Jet Interaction Generator ({\tt HIJING}) model~\cite{Wang:1991hta, Gyulassy:1994ew} is utilized. {\tt HIJING} combines a QCD inspired model for jet production with the Lund model for jet fragmentation to study jet and multi-particle production in high energy collisions. Since {\tt HIJING} does not generate genuine multi-particle correlations, it has been widely used as an ideal tool to investigate non-flow effects in the study of anisotropic flow. In the following sections, we use the {\tt toyMC}, {\tt MC-Glauber}, and {\tt HIJING} models for detailed comparisons and discussions concerning the newly studied cumulants. 

% While the {\tt AMPT} model is a hybrid model. It consists of four main stages: initial conditions from  {\tt HIJING}, partonic interactions within Zhang's Parton Cascade, hadronization modelled by a simple coalescence model~\cite{Chen:2005mr} and finally hadronic rescatterings with ART~\cite{Li:1995pra}. With the proper tuned input parameters, the {\tt AMPT} model could quantitatively or even quantitatively reproduce the particle multiplicity distributions as well as the anisotropic flow of charged hadrons in heavy-ion collisions~\cite{}. The two Monte Carlo models are used to better understand the possible non-flow effects and the behaviours of multi-particle mixed harmonic correlations from transport model, where partonic collectivity has been produced.

\section{Generic algorithm for multi-particle cumulants of azimuthal correlations}
\label{sec:ga}

The cumulant method has been used widely in high energy physics for the study of fluctuations of conserved quantities (charge, baryon, and strangeness) fluctuations to search for the QCD critical point~\cite{Luo:2017faz}, as well as in anisotropic flow studies with multi-particle correlations to eliminate non-flow contamination. It is a well established technique for evaluating the correlation between parameters of the $p.d.f.$. The implementation, however, can vary dramatically depending on how one chooses the fundamental observables. One proposal is to use particle multiplicity as the observable~\cite{DiFrancesco:2016srj} such that one determines the probability of observing a particle in a given phi window. A more popular and classical approach, however, is to utilise the azimuthal angle of the emitted particles as the fundamental observable~\cite{Borghini:2000sa} such that it reveals the probability that the particle is emitted in a specific azimuthal direction.

The mathematical formalism for evaluating cumulants is established in \cite{Kubo}. One can calculate a cumulant from moments of a distribution using the following formula:
\begin{equation}
{\rm Cum}(\{n\}) = \sum_{l=1}^n (l-1)! (-1)^{l-1} \sum_{\sum_{i=1}^l \{m_i\} = \{n\}} \prod_{i=1}^l {\rm Corr} (\{m_i\}),
\end{equation}
where $\sum_{i=1}^l \{m_i\} = \{n\}$ denotes all possible ways to split $\{n\}$ into $l$ subsets and ${\rm Corr}(\{m_i\})$ is the average value of the multiplication of all elements of $\{m_i\}$.
Multiple general algorithms for computing ${\rm Corr}(\{m_i\})$ efficiently in a single event were presented in \cite{Bilandzic:2013kga}.
A slightly simplified algorithm (with fewer user inputs) is shown here written in simple C++).

\onecolumngrid

\begin{verbatim}
complex Correlator(int* harmonic, int n, int mult = 1, int skip = 0)
 {
   int har_sum = 0;
   for (int i = 0; i<mult; ++i) har_sum += harmonic[n-1+i];
   complex c(Q(har_sum, mult));
   if (n == 1) return c;
   c *= Correlator(harmonic, n-1);
   if (n == 1+skip) return c;

   complex c2 = 0;
   int h_hold = harmonic[n-2];
   for (int counter = 0; counter <= n-2-skip; ++counter)
   {
     harmonic[n-2] = harmonic[counter];
     harmonic[counter] = h_hold;
     c2 += Correlator(harmonic, n-1, mult+1, n-2-counter);
     harmonic[counter] = harmonic[n-2];
   }
   harmonic[n-2] = h_hold;
   return c-mult*c2;
 }
\end{verbatim}

\twocolumngrid

One must provide a list of harmonics (in {\tt harmonic}) with a length of {\tt n}.
The other parameters should be left to the default values and are used internally during recursion.
Note that the code operates on complex numbers and therefore requires some class for this (here it is the class {\tt complex}).
In this code, the function for evaluating the Q-vectors ({\tt Q(n, p)}) is expected to be defined externally and should be $\sum_{k=1}^{M} w_k^p e^{i n \varphi_k}$, where $M$ is the number of particles in the event, $\varphi_k$ is the azimuthal angle of the $k^{\rm th}$ particle, $n$ is the harmonic being considered, $w_k$ is a weight given to that particle to correct for its efficiency of detection, and $p$ is a power of the weight determined by the algorithm for calculating the correlation (but can generally be as high as the order of the correlator, {\tt n}).
It was noted in \cite{Bilandzic:2013kga} that the number of terms in this function follow the Bell sequence (1, 2, 5, 15, 52, \ldots) since it requires all possible partitions of the harmonics.
This is, in fact, what is required of the cumulant itself.
With some small modifications to the routine (basically changes of constants), one can use the same method to calculate the cumulant.

\onecolumngrid

\begin{verbatim}
complex Cumulant(int* harmonic, int n, bool remove_zeros=true, int negsplit=-1,
   int mult = 1, int skip = 0)
 {
   bool remove_term = false;
   if (remove_zeros)
   {
     int har_sum = 0;
     for (int i = 0; i<mult; ++i) har_sum += harmonic[n-1+i];
     if (har_sum != 0) remove_term = true;
   }
   complex c = 0;
   if (!remove_term)
   {
     c = Corr(harmonic+(n-1), mult);
     if (n == 1) return c;
     c *= negsplit*Cumulant(harmonic, n-1, remove_zeros, negsplit-1);
   }

   int h_hold = harmonic[n-2];
   for (int counter = 0; counter <= n-2-skip; ++counter)
   {
     harmonic[n-2] = harmonic[counter];
     harmonic[counter] = h_hold;
     c += Cumulant(harmonic, n-1, remove_zeros, negsplit, mult+1, n-2-counter);
     harmonic[counter] = harmonic[n-2];
   }
   harmonic[n-2] = h_hold;
   return c;
 }
\end{verbatim}

\twocolumngrid

Again one must input an array of harmonics (in the {\tt harmonic} variable) with a length of {\tt n}.
The {\tt remove\_zeros} variable is set to true by default and removes correlations that have a residual dependence on the angular orientation of the collision and must, therefore, be zero on average over many events.
All other inputs should be left as the default values (and are used on the recursive steps).
Also note that the function {\tt Corr} is assumed to be defined externally, but is the {\it event averaged} correlation for a list of harmonics (i.e. the result of the {\tt Correlator} function averaged over many events).

The development of multi-particle cumulants of azimuthal angle correlations has been done previously in {\tt Mathematica}~\cite{Bilandzic:2012wva}.
However, that code generates a static formula that must be inserted into the computational code.
Furthermore, considering the growing number of terms in higher order cumulants, it can be tedious or even impossible to implement the code manually.
In contrast, it is rather straightforward with the generic algorithm proposed above.
The number of terms to be added in the cumulant grows quickly though following the Bell sequence where the number of terms for 6- and 8-particle cumulants are 203 and 4140, respectively.
However, this 'full cumulant' contains terms that have averages for single harmonics which must become zero due to the isotropic distribution over many events.
If one excludes these terms, the number of remaining terms (which could be non-zero depending on which harmonics are chosen) reduces considerably.
For a 6-particle cumulant one gets 41 terms while for 8-particle cumulant there are 715 terms.
A further reduction of terms is achieved when one calculates the cumulant with specific harmonics and removes terms involving correlations where the sum of harmonics is not 0 (since these must also be zero due to the isotropic distribution of particles over many events).
This is accomplished by leaving the {\tt remove\_zeros} parameter as {\tt true} in the code above.
Finally, to obtain an optimised formula, when harmonics are repeated, one can reduce the number of terms by grouping identical ones together.
It should be noted that the generic algorithm for the cumulant above does not do this optimization, though.

Based on the above generic algorithm, we can obtain all possible 4-particle cumulants (without terms involving only a single particle which vanish after averaging over many events due to symmetry):
\begin{eqnarray}
&& Cum(n_1, n_2, n_3, n_4)  = \left< \left< e^{ i \, (n_1\varphi_1 + n_2 \varphi_2 + n_3\varphi_3 + n_4\varphi_4 )}\right>\right>_c  \nonumber\\
&=& \left< \left< e^{ i \, (n_1\varphi_1 + n_2 \varphi_2 + n_3\varphi_3 + n_4\varphi_4 )}\right>\right>  \nonumber\\
&~& -   \left< \left< e^{i \,(n_1\varphi_1 + n_2 \varphi_2  )}\right>\right> \, \left< \left< e^{i \,(n_3\varphi_3 + n_4 \varphi_4  )}\right>\right>   \nonumber\\
&~& -  \left< \left< e^{i \, (n_1\varphi_1 + n_3 \varphi_3  )}\right>\right> \, \left< \left< e^{i \,(n_2\varphi_2 + n_4 \varphi_4  )}\right>\right> \nonumber\\
&~& - \left< \left< e^{ i \,(n_1\varphi_1 + n_4 \varphi_4 )}\right>\right> \, \left< \left< e^{i \, (n_2\varphi_2 + n_3 \varphi_3 ) }\right>\right>  
\label{eq:4pc_all}
\end{eqnarray}
Equation~(\ref{eq:4pc_all}) is general for any set of $(n_1,n_2,n_3,n_4)$. Once specific harmonics are chosen, one can further test if $n_1 + n_2 + n_3 + n_4 = 0$, since not satisfying this would again make the term vanish. Therefore, if $n_1=n, n_2=n, n_3=-n, n_4=-n$, Eq.~(\ref{eq:4pc_all}) will be modified to:
\begin{eqnarray}
 Cum(n, n, -n, -n) &=&  \left< \left< e^{i \, n (\varphi_1 + \varphi_2 - \varphi_3 - \varphi_4 ) }\right>\right>_c  \nonumber\\
&=& \left< \left< e^{i \, n (\varphi_1 + \varphi_2 - \varphi_3 - \varphi_4 ) } \right>\right>   \\ & ~& -   2 \left< \left< e^{ i\, n (\varphi_1 - \varphi_2  )} \right>\right>^2   \nonumber\\
&=& \left< v_{n}^{4} \right> - 2 \left< v_{n}^{2} \right>^2
\label{eq:4pc_cn4}
\end{eqnarray}
This is the standard four-particle cumulant of a single harmonic which is usually called $c_{n}\{4\}$. In the case of $n_1=k$, $n_2=l$, $n_3=-k$, and $n_4=-l$ where $k \neq l$, Eq.~(\ref{eq:4pc_all}) can be written as:
\begin{eqnarray}
 Cum(k, l, -k, -l)   &=&  \left< \left< e^{ i \, (k\varphi_1 + l\varphi_2 - k\varphi_3 - l\varphi_4 ) } \right>\right>_c  \nonumber\\
&=& \left< \left< e^{i\, (k\varphi_1 + l\varphi_2 - k\varphi_3 - l\varphi_4 ) } \right>\right> \nonumber\\
&& -   \left< \left< e^{i \, k (\varphi_1 - \varphi_2  )} \right>\right> \, \left< \left< e^{i \, l (\varphi_1 - \varphi_2  )} \right>\right>   \nonumber\\
&=& \left< v_{k}^{2} \, v_{l}^{2} \right> - \left< v_{k}^{2} \right> \,  \left< v_{l}^{2} \right>
\label{eq:4pc_SC}
\end{eqnarray}
This type of 4-particle cumulant of mixed harmonics is also known as the symmetric cumulant, $SC(k,l)$~\cite{Bilandzic:2013kga}.

\section{higher order cumulant of single harmonic}
\label{sec:single}

It is well known that anisotropic flow coefficients fluctuate event-by-event even in a narrow centrality bin. Besides the event-by-event Bayesian unfolding approach~\cite{Aad:2013xma} whose implementation is complex, multi-particle cumulants are a widely used approach to quantify the fluctuations of $v_n$ due to the fact that various cumulants have unique sensitivity to the moments of the $P(v_n)$. The relevant studies~\cite{Acharya:2018lmh, Aad:2014vba, Zhou:2015eya} have shown that $P(v_n)$ is better described by the Elliptic-Power function~\cite{Yan:2013laa, Yan:2014afa} relative to the Bessel-Gaussian function~\cite{Voloshin:2007pc}, which was commonly used in previous fluctuation studies. So far, cumulants only up to 8-particles have been studied. higher order cumulants could bring additional sensitivity to $P(v_n)$ but have not been measured due to the difficulties both in the construction of multi-particle cumulants as well as in the implementations of the multi-particle correlations which are the fundamental element of a multi-particle cumulant. For the multi-particle cumulants involving a single harmonic we have from the previous studies:
\begin{eqnarray}
v_{n}\{2\}^2 &= & \langle v_{n}^{2} \rangle, \nonumber\\ 
v_{n}\{4\}^4 &= & - \bigg( \langle v_{n}^{4} \rangle - 2\langle v_{n}^{2} \rangle^{2} \bigg), \nonumber\\ 
v_{n}\{6\}^6 &= & \frac{1}{4}  \bigg( \langle v_{n}^{6} \rangle - 9\langle v_{n}^{4} \rangle \langle v_{n}^{2} \rangle + 12\langle v_{n}^{2} \rangle^{3}  \bigg) , \nonumber\\ 
v_{n}\{8\}^8 &= &   - \frac{1}{33}  \bigg( \langle v_{n}^{8} \rangle  - 16 \langle  v_{n}^{6} \rangle \langle v_{n}^{2} \rangle - 18 \langle v_{n}^{4} \rangle^{2} \nonumber\\ 
&& + 144 \langle v_{n}^{4} \rangle \langle v_{n}^{2}  \rangle^{2}  - 144 \langle v_{n}^{2} \rangle^{4} \bigg) . 
\end{eqnarray}

All of these have been studied systematically both in experimental measurements and theoretical calculations. Now using the newly proposed generic algorithm of multi-particle cumulants introduced in section~\ref{sec:ga}, we can easily construct any multi-particle cumulant that we are interested in. For the higher order cumulants of a single harmonic involving more than 8-particles we have, after grouping repeated terms together, the following formulas\footnote{These formulas were independently derived in a Master's thesis at the Niels Bohr Institute by Troels Krogsb{\o}ll.}:
\begin{eqnarray}
v_{n}\{10\}^{10} &=& \frac{1}{456} \,  \bigg( \left< v_{n}^{10} \right>  -  25\, \left<  v_n^8 \right> \,\left<  v_n^2 \right> - 100\left<  v_n^6 \right> \, \left< v_n^4 \right>   \nonumber\\ 
 && +400 \, \left<  v_n^6 \right> \, \left<  v_n^2 \right>^{2}  + 900\, \left<  v_n^4 \right>^{2} \, \left< v_n^2 \right>  \nonumber\\ 
& & -3600 \, \left<  v_n^4 \right> \, \left< v_n^2 \right>^{3}  + 2880 \, \left<  v_n^2 \right>^5  \bigg)
\end{eqnarray}

\begin{eqnarray}
v_{n}\{12\}^{12} &=& - \frac{1}{9460} \bigg( \left< v_{n}^{12} \right>  -  36\left<  v_n^{10} \right> \,\left<  v_n^2 \right> - 225\left<  v_n^8 \right> \, \left< v_n^4 \right>   \nonumber\\ 
 && +900 \, \left<  v_n^8 \right> \, \left<  v_n^2 \right>^{2}  - 200\, \left<  v_n^6 \right>^{2} \nonumber\\ 
& & +7200 \, \left<  v_n^6 \right> \, \left< v_n^4 \right> \, \left< v_n^2 \right>  -14400 \, \left<  v_n^6 \right> \, \left<  v_n^2 \right>^3  \nonumber\\ 
& & +2700 \, \left<  v_n^4 \right>^3  -48600 \, \left<  v_n^4 \right>^2 \, \left<  v_n^2 \right>^2  \nonumber\\ 
& & +129600 \, \left<  v_n^4 \right> \, \left<  v_n^2 \right>^4  - 86400 \, \left<  v_n^2 \right>^6   \bigg) 
\end{eqnarray}

\begin{eqnarray}
v_{n}\{14\}^{14} &=& \frac{1}{274800} \bigg( \left< v_{n}^{14} \right>  -  49\, \left<  v_n^{12} \right> \,\left<  v_n^2 \right>   \nonumber\\
 && - 441\, \left<  v_n^{10} \right> \, \left< v_n^4 \right> + 1764\, \left<  v_n^{10} \right> \, \left<  v_n^2 \right>^{2}   \nonumber\\
 && - 1225\, \left<  v_n^8 \right> \, \left<  v_n^6 \right> + 22050 \, \left<  v_n^8 \right> \, \left< v_n^4 \right> \, \left< v_n^2 \right>   \nonumber\\
 && - 44100 \, \left<  v_n^8 \right> \, \left<  v_n^2 \right>^3 + 19600 \, \left<  v_n^6 \right>^2 \, \left< v_n^2 \right>   \nonumber\\
 && + 44100 \, \left<  v_n^6 \right> \, \left<  v_n^4 \right>^2 - 529200 \, \left<  v_n^6 \right> \, \left<  v_n^4 \right> \, \left<  v_n^2 \right>^2   \nonumber\\
 && + 705600 \, \left<  v_n^6 \right> \, \left<  v_n^2 \right>^4 - 396900 \, \left<  v_n^4 \right>^3 \, \left<  v_n^2 \right>   \nonumber\\
 && + 3175200 \, \left<  v_n^4 \right>^2 \, \left<  v_n^2 \right>^3 - 6350400 \, \left<  v_n^4 \right> \, \left<  v_n^2 \right>^5   \nonumber\\
 && + 3628800 \left<  v_n^2 \right>^7  \bigg)
\end{eqnarray}

\begin{eqnarray}
v_{n}\{16\}^{16} &=& - \frac{1}{10643745} \bigg( \left< v_{n}^{16} \right>  -  64\, \left<  v_n^{14} \right> \,\left<  v_n^2 \right>   \nonumber\\ 
 && - 784\left<  v_n^{12} \right> \, \left< v_n^4 \right>  +3136 \, \left<  v_n^{12} \right> \, \left<  v_n^2 \right>^{2} \nonumber\\ 
 && - 3136\, \left<  v_n^{10} \right> \, \left<  v_n^6 \right>   +56448 \, \left<  v_n^{10} \right> \, \left< v_n^4 \right> \, \left< v_n^2 \right>  \nonumber\\ 
 && - 112896 \, \left<  v_n^{10} \right> \, \left<  v_n^2 \right>^3  -2450 \, \left<  v_n^8 \right>^2 \nonumber\\ 
 && + 156800 \, \left<  v_n^8 \right> \, \left<  v_n^6 \right>\, \left<  v_n^2 \right> +176400 \, \left<  v_n^8 \right> \, \left<  v_n^4 \right>^2 \nonumber\\ 
 && - 2116800 \, \left<  v_n^8 \right> \, \left<  v_n^4 \right>  \, \left<  v_n^2 \right>^2   \nonumber\\
 && + 2822400 \, \left<  v_n^8 \right> \left<  v_n^2 \right>^4 + 313600 \, \left<  v_n^6 \right>^2 \, \left<  v_n^4 \right>   \nonumber\\
 && - 1881600 \, \left<  v_n^6 \right>^2 \, \left<  v_n^2 \right>^2   \nonumber\\
 && - 8467200 \left<  v_n^6 \right>  \, \left<  v_n^4 \right>^2 \, \left<  v_n^2 \right>   \nonumber\\
 && + 45158400 \, \left<  v_n^6 \right> \, \left<  v_n^4 \right>  \, \left<  v_n^2 \right>^3   \nonumber\\
 && - 45158400 \left<  v_n^6 \right>  \, \left<  v_n^2 \right>^5  - 1587600 \, \left<  v_n^4 \right>^4   \nonumber\\
 && + 50803200 \left<  v_n^4 \right>^3  \, \left<  v_n^2 \right>^2   \nonumber\\
 && - 254016000 \, \left<  v_n^4 \right>^2  \, \left<  v_n^2 \right>^4   \nonumber\\
 && + 406425600 \left<  v_n^4 \right>  \, \left<  v_n^2 \right>^6 - 203212800 \left<  v_n^2 \right>^8 \bigg) \nonumber\\
\end{eqnarray}

\begin{figure}[htbp!] 
\begin{center}
\includegraphics[width=0.98\linewidth]{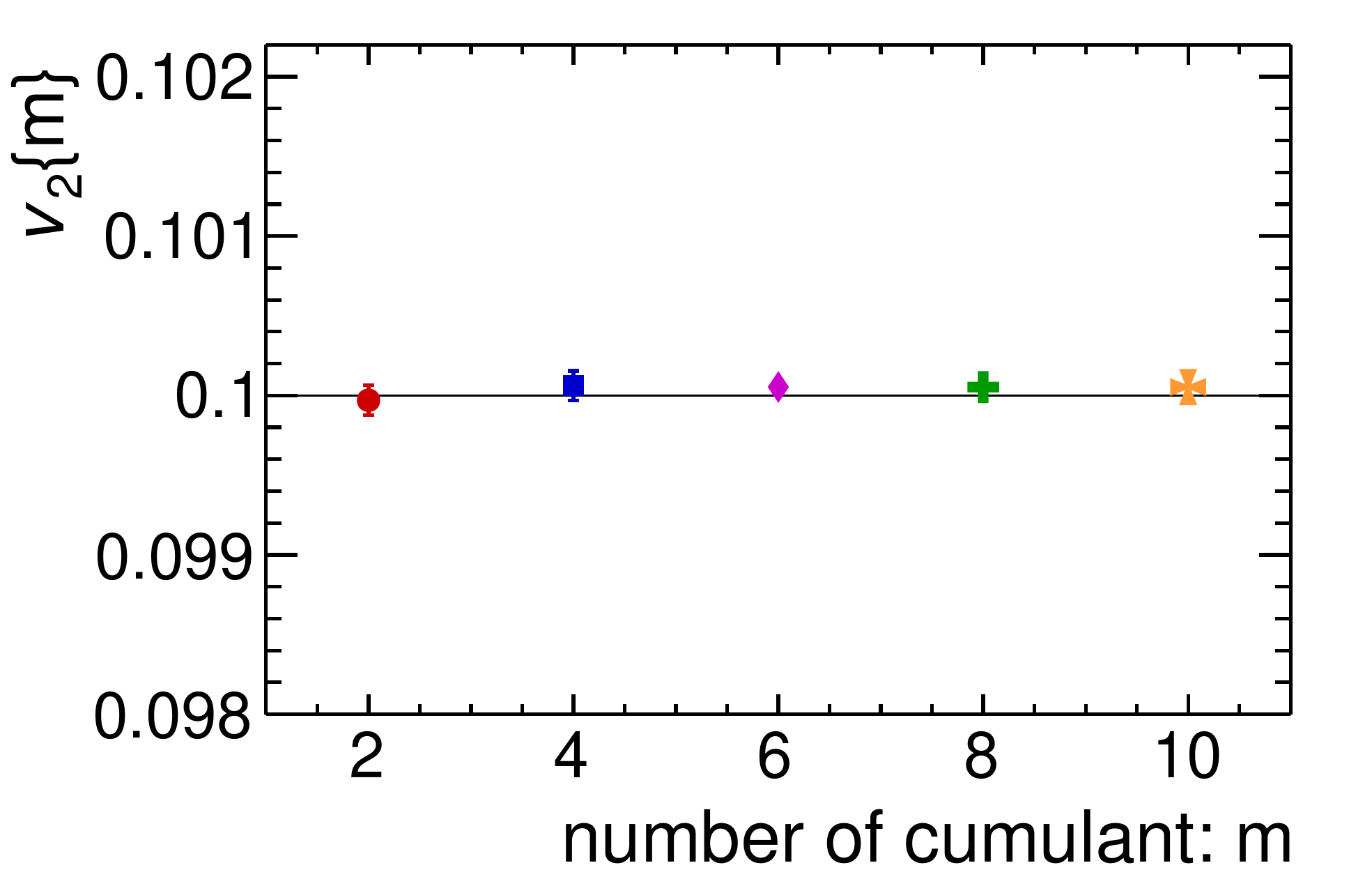}
\caption{(Color online) Multi-particle cumulants of $v_2$ from toy Monte Carlo simulations. The input value of $v_2$ is 0.1. All cumulants retrieve this value with high precision.}
\label{fig:cum_toy} 
\end{center}
\end{figure}

\begin{figure}[htbp!]
\centering
\includegraphics[width=0.98\linewidth]{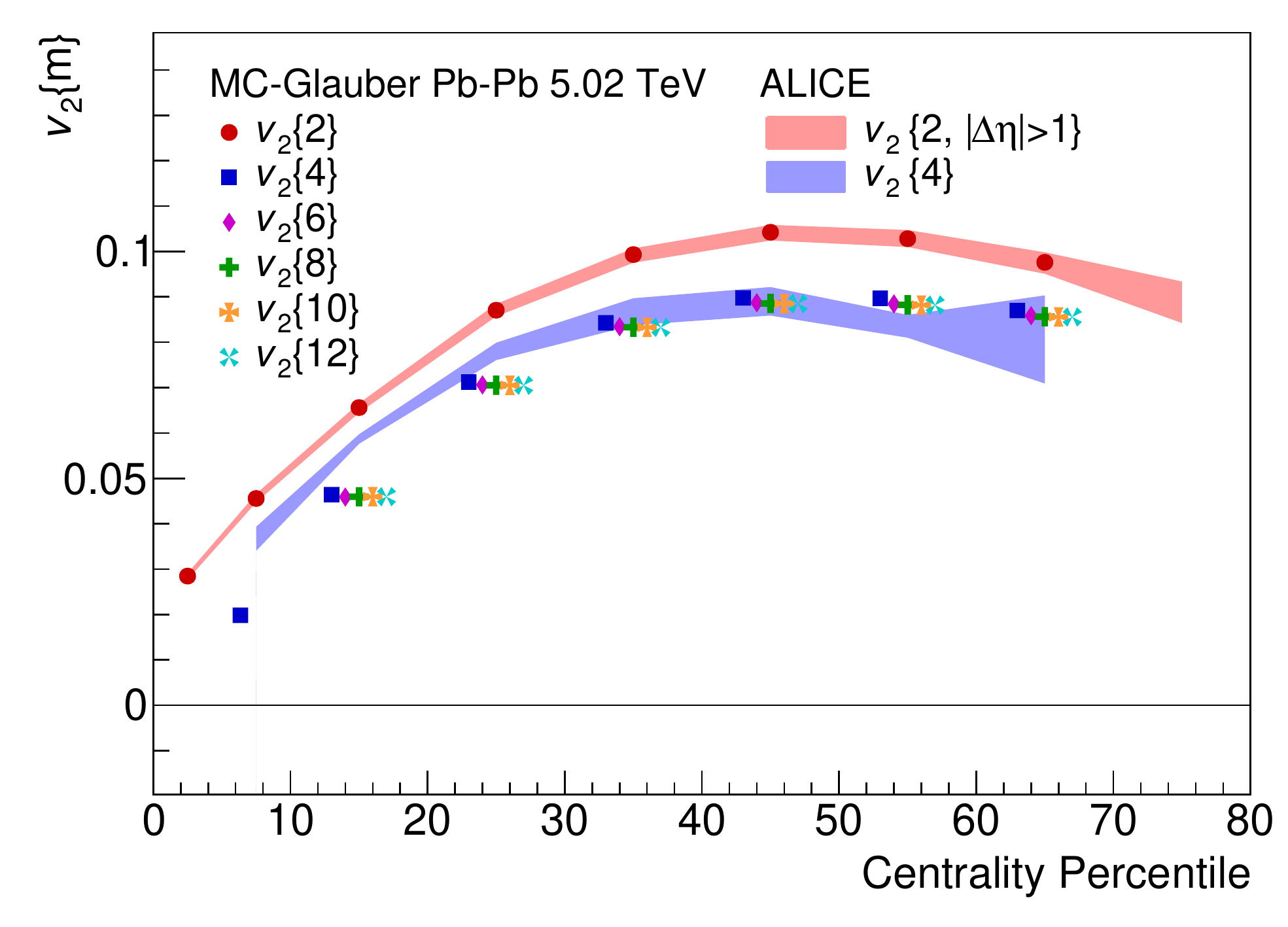}
\caption{(Color online) Multi-particle cumulants of $v_2$ from the {\tt MC-Glauber} model using a scaled value of the $\kappa_2$ parameter. The ALICE data presented here are from \cite{Adam:2016izf}.}
\label{fig:cum_MCG} 
\end{figure}

\begin{figure*}[htb]
\centering
\includegraphics[width=0.98\linewidth]{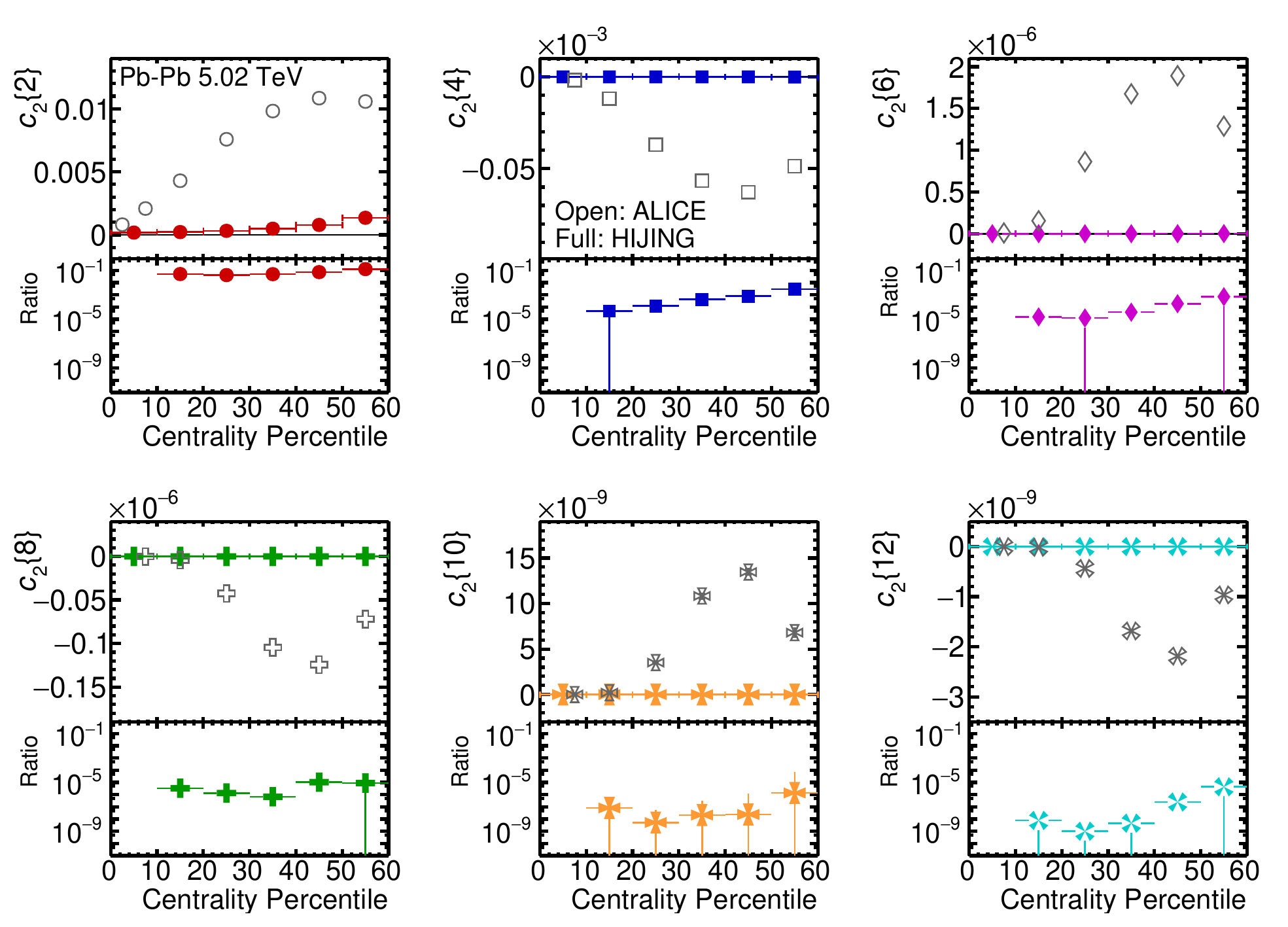}
\caption{(Color online) Centrality dependence of multi-particle cumulants of $v_2$ for all charged hadrons in Pb--Pb collisions at 5.02 TeV. Calculations from {\tt HIJING} (full markers) are compared to the ALICE data (open markers) from~\cite{Adam:2016izf}. } 
\label{fig:cum}
\end{figure*}

To validate the equations of multi-particle cumulants from the generic algorithm we perform the {\tt toyMC} study shown in Fig.~\ref{fig:cum_toy}. As introduced in Section~\ref{sec:model}, the input $v_2$ and $v_3$ are fixed to 0.1 and 0.05, respectively, with no event-by-event flow fluctuations. Thus, one expects that $v_{2}\{2\} = v_{2}\{4\} = v_{2}\{6\} = v_{2}\{8\} = v_{2}\{10\} =  ... = v_{2}\{m\} = v_{2}^{\rm input} $. Figure~\ref{fig:cum_toy} shows that this is indeed what is seen as there is excellent agreement between the multi-particle cumulants of $v_2$ and the input $v_2$ value of 0.1. This {\tt toyMC} study therefore validates the generic algorithm and its resulting formulas of multi-particle cumulants of a single harmonic. 

In order to study event-by-event fluctuations of $v_n$ to high orders, we employ {\tt MC-Glauber} simulations. Assuming that $v_2 = \kappa_2 \cdot \varepsilon_2$, we can firstly calculate $\varepsilon_2$ for every single event and then scale it to obtain $v_2$ for that event. Here $\kappa_2$ is calculated from the ratio of ALICE measurements of $v_2\{2, |\Delta \eta| > 1.0\}$ in Ref.~\cite{Adam:2016izf} and $\varepsilon_2\{2\}$ from {\tt MC-Glauber}. The value of $\kappa_2$ depends on the collision centrality. In the end, we obtain the multi-particle cumulants of $v_2$ from an {\tt MC-Glauber} simulation, which are shown in Fig.~\ref{fig:cum_MCG}. It is known that 2- and multi-particle cumulants have different sensitivities to flow fluctuations, which are positive for a 2-particle cumulant and negative for multi-particle cumulants. Here for the first time, we study the effect of flow fluctuations on very high order cumulants. It can be seen that $v_{2}\{2\} > v_{2}\{4\} = v_{2}\{6\} = v_{2}\{8\} = v_{2}\{10\} = v_{2}\{12\}$. The first 10- and 12-particle cumulants studies show that these two higher order cumulants also have a negative sensitivity to flow fluctuations, just as 4-, 6-, and 8-particle cumulants have. It should be emphasised that the difference between high order cumulants of $v_2$ has already been found to be below 0.5\%. Considering the current precision of experimental measurements, especially the systematical uncertainty which is usually not smaller than 0.5\%, it will be extremely challenging to probe the possible tiny differences between even higher order cumulants measured in heavy-ion experiments. However, higher order cumulants with more than 8-particles will be very useful for investigating collective flow in small collision systems and constraining its origin, whether it is due to pure final state effects, e.g. the hydrodynamic response to the initial geometry and its fluctuations, or from the effect of initial momentum correlations. The evidence for flow in small systems is partially based on $v_2$ from multi-particle cumulants being compatible within large statistical and systematical uncertainties. However, some non-negligible non-flow contamination still remains, and it could bias the data and theory comparisons, thus resulting in an incorrect physics conclusion.

In order to avoid the effect of non-flow, the sub-event method of multi-particle cumulants was developed and proved to be effective~\cite{Jia:2017hbm,Huo:2017nms}. The application of 10-, 12-, and even 14-particle cumulants, on the other hand, should enable a new possibility to further reduce the effect of non-flow contamination. This is tested here by studying multi-particle cumulants using {\tt HIJING} simulations. Considering the fact that 10- and 12-particle cumulants of $v_2$ were never measured before, we assume that $v_{2}\{10\} = v_{2}\{12\} = v_{2}\{4\}$ and then we have $c_{2}\{10\} = 456 \cdot v_{2}\{10\}^{10}$ and $c_{2}\{12\} = -9460 \cdot v_{2}\{12\}^{12}$.
The ratios of calculations from {\tt HIJING} simulations to experimental data are presented in Fig.~\ref{fig:cum}. In the case that the multi-particle cumulant $c_{2}\{m\}$ has an opposite sign relative to data, the absolute value of the ratio is plotted. As can be seen in Fig.~\ref{fig:cum}, the non-flow contamination becomes smaller when one uses higher order cumulants of $c_2$. Thus, one can expect that measurements of high order cumulants of a single harmonic, $v_n$, in the coming high luminosity LHC Run 3 program will provide more reliable data and theory comparisons, in addition to the existing body of work and will eventually further our understanding of the origin of collective flow in small collision systems.

\section{Correlations between different moments of flow coefficient}
\label{sec:mixed}

The idea of using multi-particle mixed harmonic cumulants to study the flow vector correlations or in general probe the joint $p.d.f.$ has existed for a while (e.g. in Ref.~\cite{Zhou:2015eya}). However, the implementation of arbitrary order multi-particle mixed harmonics cumulants was not easily achieved before the development of the generic algorithm introduced above. Among many mixed harmonic cumulants which could involve not only $v_n$, $v_m$ but also $\Psi_n$ and $\Psi_m$, there is a series of cumulants which contain only correlations between different order flow coefficients without flow symmetry planes. In particular, the symmetric cumulant, defined as $SC(m,n) = \left<v_{m}^2v_{n}^2\right>-\left<v_{m}^2\right>\left<v_{n}^2\right>$, was proposed to quantify the correlations between the second moments of $v_n$ and $v_m$~\cite{Bilandzic:2013kga}. In a narrow phase space where flow vector fluctuations are negligible, $SC(m,n)$ is independent of the symmetry planes $\Psi_m$ and $\Psi_n$ and is expected to be less sensitive to non-flow correlations. This was confirmed by the $SC(m,n)$ calculation using {\tt HIJING} model~\cite{Wang:1991hta,Gyulassy:1994ew}, which does not include anisotropic collectivity but e.g. azimuthal correlations due to jet production. (Although it is also noticed that for the small collision systems like proton-lead and proton-proton collisions there is still significant non-flow contamination, thus, the sub-event method using 4-particle cumulants has to be applied~\cite{Huo:2017nms}). Therefore, it is believed that in the heavy-ion collisions, $SC(m,n)$ is non-zero if there are non-trivial (anti-)correlations between different flow coefficients. It should, however, be emphasized that $SC(m,n)$ measures only the correlation between $v_n^2$ and $v_m^2$ and one can gain more information about correlations between different moments by looking at higher order correlations. 

Here we propose to investigate these higher order multi-particle mixed harmonic cumulants further. We would like to go beyond the completely symmetric cumulants to include higher order mixed-harmonic cumulants that are independent of the event plane. To quantify the correlation between the $2k^{th}$ flow coefficient of $v_m$ and the $2l^{th}$ moment of $v_n$, or more specifically the correlation between $v_m^{2 k}$ and $v_n^{2 l}$, we use the notation $MHC(v_m^{2 k}, v_n^{2 l})$. This notation can be expanded to include more than two harmonics, but one must note that to be independent of the event plane there must not exist the set of harmonics used should not be divisible into subsets of harmonics that are equal in sum unless the subset contains a single harmonic. i.e. for three harmonics, $m$, $n$, and $p$, it should not be that $m+n=p$. See a discussion about this in Appendix~\ref{appb}. For correlations between the second moments of two harmonics, $SC(m,n)$ is identical to $MHC(v_m^2, v_n^2)$. Here we list several typical mixed harmonic cumulants which involve 6- and 8-particles.

%In a more general case, to quantify the correlation between the $k^{th}$ flow moment of $v_m$ and the $l^{th}$ moment of $v_n$, or more specifically the correlation between $v_m^k$ and $v_n^l$, we use the notation $MHC(v_m^k, v_n^l)$~\footnote{For convinence, we will use $MHC$ in the following text}. Obviously, for correlations between the second moments of two harmonics, $SC(m,n)$ is identical to $MHC(v_m^2, v_n^2)$. Here we list several typical mixed harmonic cumulants which involve 6- and 8-particles. 

For 6-particle cumulants, one can quantify the correlation between $v_2^4$ and $v_3^2$ as:
\begin{eqnarray}
MHC(v_2^4, v_3^2) &=&  \left< \left< e^{ i \,(2\varphi_1 + 2 \varphi_2 + 3\varphi_3 - 2\varphi_4  - 2\varphi_5  - 3\varphi_6  ) }\right>\right>_c \nonumber\\
&=&  \left<  v_2^4 \, v_3^2 \right> -  4 \,\left<  v_2^2 \, v_3^2 \right> \, \left<  v_2^2 \right> -  \left<  v_2^4 \right> \, \left<  v_3^2 \right>  \nonumber\\
& & + 4 \,\left<  v_2^2 \right>^2  \left<  v_3^2 \right> 
\end{eqnarray}
Here the lower order (including both 2- and 4-particle) correlations have been subtracted from the 6-particle correlation. Therefore, if there is no correlations between $v_2^4$ and $v_3^2$, then $MHC(v_2^4, v_3^2)$ is expected to be consistent with zero. Similarly, one can obtain $MHC(v_2^2, v_3^4)$:
\begin{eqnarray}
MHC(v_2^2, v_3^4) &=&  \left< \left< e^{ i \,(2\varphi_1 + 3 \varphi_2 + 3\varphi_3 - 2\varphi_4  - 3\varphi_5  - 3\varphi_6  ) }\right>\right>_c \nonumber\\
&=&  \left<  v_2^2 \, v_3^4 \right> -  4 \,\left<  v_2^2 \, v_3^2 \right> \, \left<  v_3^2 \right> -  \left<  v_2^2 \right> \, \left<  v_3^4 \right>  \nonumber\\
& & + 4 \,\left<  v_2^2 \right>  \left<  v_3^2 \right>^2 
\end{eqnarray}

If we continue to 8-particle cumulants, one can study the correlation between $v_2^6$ and $v_3^2$ as:
\begin{eqnarray}
MHC(v_2^6, v_3^2) &=&  \left< \left< e^{ i \,(2\varphi_1 + 2 \varphi_2 + 2\varphi_3 + 3 \varphi_4  - 2\varphi_5  - 2\varphi_6  - 2\varphi_7  - 3\varphi_8) }\right>\right>_c \nonumber\\
&=& \left<  v_2^6 \, v_3^2 \right>  -  9\, \left<  v_2^4 \, v_3^2 \right> \,\left<  v_2^2 \right> - \left<  v_2^6 \right> \, \left< v_3^2 \right>   \nonumber\\ 
 &&- 9 \, \left<  v_2^4 \right> \, \left<  v_2^2 \, v_3^2 \right>  - 36\, \left<  v_2^2 \right>^3 \, \left< v_3^2 \right>  \nonumber\\ 
& & + 18 \, \left<  v_2^2 \right> \, \left< v_3^2 \right> \, \left<  v_2^4 \right> + 36 \, \left<  v_2^2 \right>^2\,\left<  v_2^2 \, v_3^2 \right>  \nonumber\\ 
\end{eqnarray}
Analogously, one can get $MHC(v_2^2, v_3^6)$ by swapping $v_2$ and $v_3$ in the equation above.
\begin{eqnarray}
MHC(v_2^2, v_3^6) 
&=&  \left< \left< e^{ i \,(2\varphi_1 + 3 \varphi_2 + 3\varphi_3 + 3 \varphi_4  - 2\varphi_5  - 3\varphi_6  - 3\varphi_7  - 3\varphi_8) }\right>\right>_c \nonumber\\
&=& \left<  v_2^2 \, v_3^6 \right>  -  9\, \left<  v_2^2 \, v_3^4 \right> \,\left<  v_3^2 \right> - \left<  v_3^6 \right> \, \left< v_2^2 \right>  \nonumber\\ 
&& - 9 \, \left<  v_3^4 \right> \, \left<  v_2^2 \, v_3^2 \right>  - 36\, \left<  v_2^2 \right> \, \left< v_3^2 \right>^3  \nonumber\\ 
& & + 18 \, \left<  v_2^2 \right> \, \left< v_3^2 \right> \, \left< v_3^4 \right> + 36 \, \left<  v_3^2 \right>^2\,\left<  v_2^2 \, v_3^2 \right>    \nonumber\\ 
&& 
\end{eqnarray}
Finally, for an 8-particle cumulant with $v_2^4$ and $v_3^4$, one gets:
\begin{eqnarray}
MHC(v_2^4, v_3^4) &=&  \left< \left< e^{ i \,(2\varphi_1 + 2 \varphi_2 + 3\varphi_3 + 3 \varphi_4  - 2\varphi_5  - 2\varphi_6  - 3\varphi_7  - 3\varphi_8) }\right>\right>_c \nonumber\\
&=&  \left< v_2^4 \, v_3^4 \right> - 4 \,  \left< v_2^4 \, v_3^2 \right>  \left< v_3^2 \right>   \nonumber\\ 
&& -  4 \, \left< v_2^2 \, v_3^4 \right> \,  \left<  v_2^2\right> -  \left<  v_2^4 \right> \,  \left< v_3^4 \right>   \nonumber\\ 
&& - 8 \,  \left< v_2^2 \, v_3^2 \right>^{2}  - 24\,  \left<  v_2^2 \right>^2 \, \left<  v_3^2 \right>^2 \nonumber \\
&& + 4\,  \left< v_2^2 \right>^2 \, \left< v_3^4 \right>  + 4 \, \left<  v_2^4 \right> \,  \left< v_3^2 \right>^2 \nonumber \\
&& + 32 \,  \left< v_2^2 \right> \, \left< v_3^2 \right> \,  \left< v_2^2 \, v_3^2\right> 
\end{eqnarray}

One can also go beyond studying multi-particle cumulants involving only two different harmonics by constructing cumulants with three or more flow harmonics. For instance, the correlation between $v_2^2$, $v_3^2$ and $v_4^2$ can be quantified as:
\begin{eqnarray}
MHC(v_2^2, v_3^2, v_4^2) &=&  \left< \left< e^{ i \,(2\varphi_1 + 3 \varphi_2 + 4\varphi_3 - 2\varphi_4  - 3\varphi_5  - 4\varphi_6  ) }\right>\right>_c \nonumber\\
&=& \left<  v_2^2 \, v_3^2 \, v_4^2 \right> -  \left<  v_2^2 \, v_3^2 \right> \, \left<  v_4^2 \right> -  \left<  v_2^2 \, v_4^2 \right> \, \left<  v_3^2 \right>  \nonumber\\ 
&& -  \left<  v_3^2 \, v_4^2 \right> \, \left<  v_2^2 \right>   + 2 \,\left<  v_2^2 \right> \, \left<  v_3^2 \right> \, \left<  v_4^2 \right> 
\label{eq:MHC234}
\end{eqnarray}
Similarly, for the correlation between $v_3^2$, $v_4^2$ and $v_5^2$, one gets:
\begin{eqnarray}
MHC(v_3^2, v_4^2, v_5^2)  &=&  \left< \left< e^{ i \,(3\varphi_1 + 4 \varphi_2 + 5\varphi_3 - 3\varphi_4  - 4\varphi_5  - 5\varphi_6  ) }\right>\right>_c \nonumber\\
&=& \left<  v_3^2 \, v_4^2 \, v_5^2 \right> -  \left<  v_3^2 \, v_4^2 \right> \, \left<  v_5^2 \right> -  \left<  v_3^2 \, v_5^2 \right> \, \left<  v_4^2 \right>  \nonumber\\ 
&& -  \left<  v_4^2 \, v_5^2 \right> \, \left<  v_3^2 \right>   + 2 \,\left<  v_3^2 \right> \, \left<  v_4^2 \right> \, \left<  v_5^2 \right> 
\label{eq:MHC345}
\end{eqnarray}

In addition, one can define the normalized $MHC$, denoted as $nMHC$, which can be expressed as:
\begin{equation}
nMHC(v_m^k,v_n^l)=\frac{MHC(v_m^k,v_n^l)}{\left\langle v_m^k\right\rangle\left\langle v_n^l\right\rangle}
\label{nMHC}
\end{equation}
Here $nMHC(v_m^k,v_n^l)$ eliminates the dependence on the magnitudes of $v_m$ and $v_n$, and therefore can be used for quantitative comparison of genuine correlations between experimental data and model calculations.

Similarly, for $MHC$ invoing three harmonics, we can have the corresponding $nMHC$ as:
\begin{equation}
nMHC(v_m^k,v_n^l, v_p^q)=\frac{MHC(v_m^k,v_n^l, v_p^q)}{\left\langle v_m^k\right\rangle \, \left\langle v_n^l\right\rangle \, \left\langle v_p^q\right\rangle}
\label{nMHCH}
\end{equation}

\begin{figure}[htbp!]
\begin{center}
\includegraphics[width=0.8\linewidth]{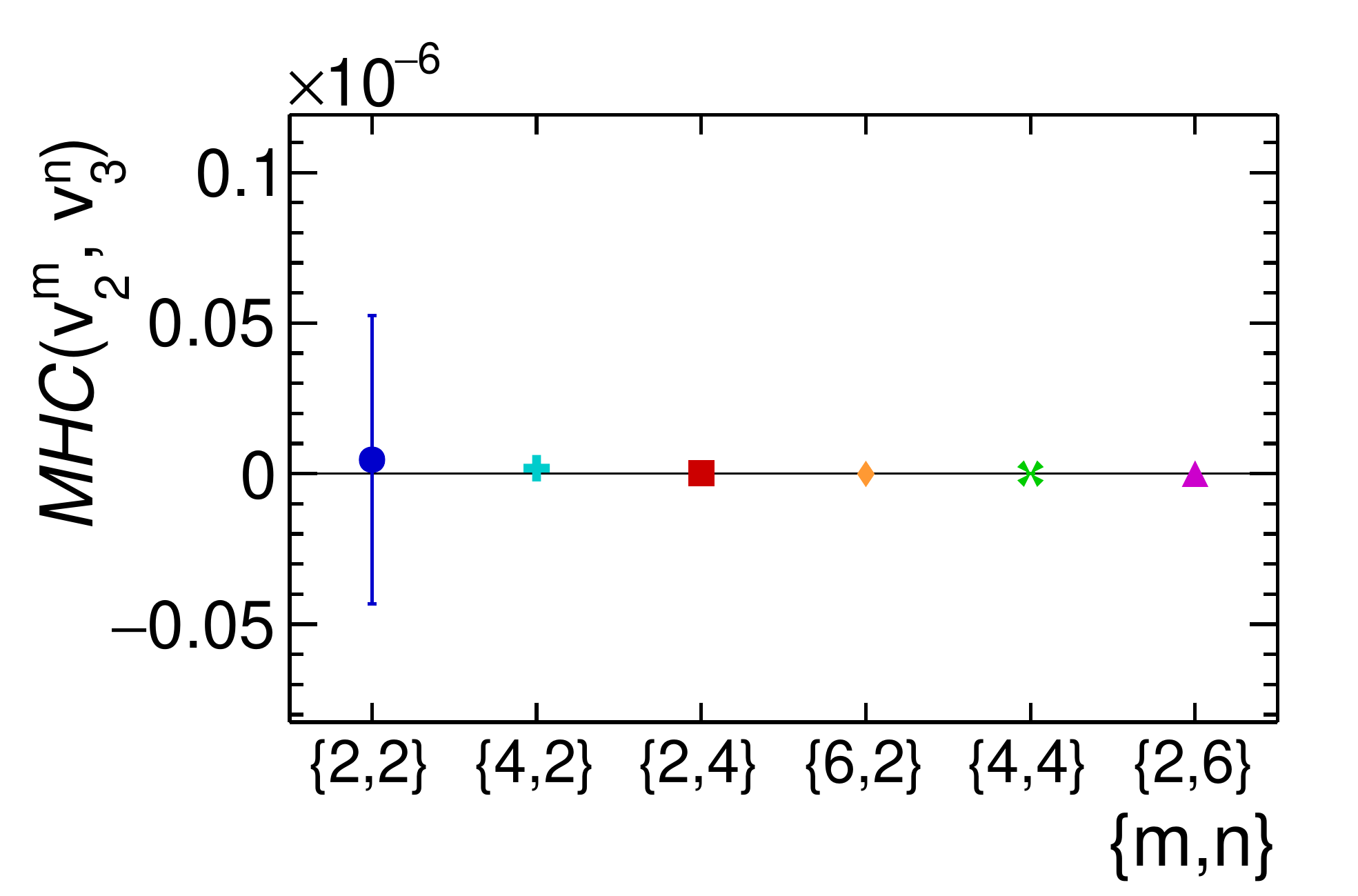}
\caption{(Color online) Multi-particle mixed-harmonic cumulants ($MHC$) from {\tt toyMC} simulations. All MHC return zero as there was no correlation put in the {\tt toyMC} simulation.}
\label{fig:MHC_toyMC} 
\end{center}
\end{figure}

The {\tt toyMC} model is first used to validate the generic algorithm and the corresponding multi-particle cumulants for mixed harmonics. The results are shown in Fig.~\ref{fig:MHC_toyMC} where it is seen that all the results are consistent with zero. This is expected due to the fact that both $v_2$ and $v_3$ have fixed values for every single event, and there is no correlation between them. %This {\tt toyMC} study presented here shows that the generic algorithm and the corresponding multi-particle cumulants for mixed harmonics do not have a trivial mistake in the implementation. 

\begin{figure}[htbp!]
\begin{center}
\includegraphics[width=0.9\linewidth]{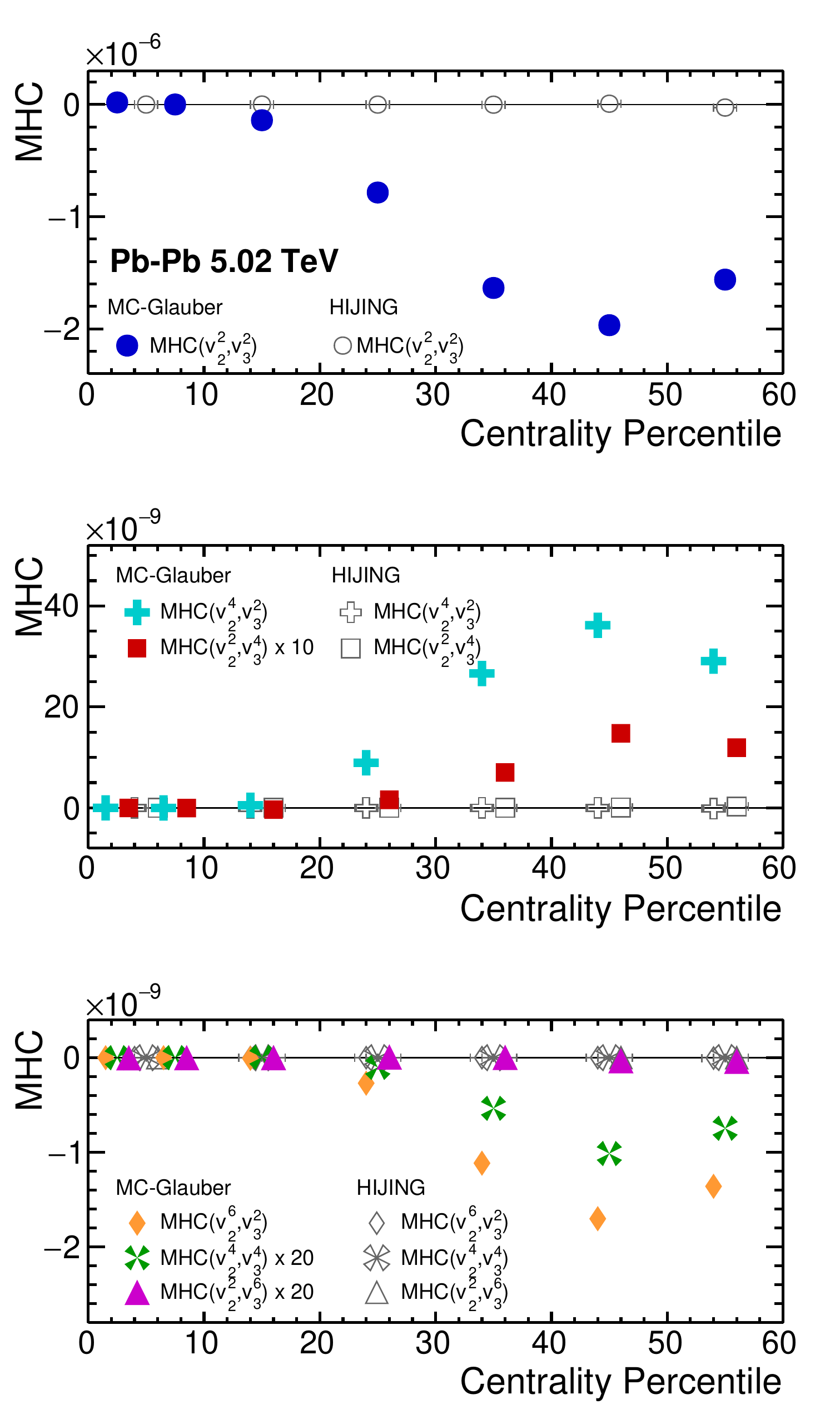}
\caption{(Color online) Mixed-harmonic cumulants ($MHC$) from the {\tt MC-Glauber} and {\tt HIJING} models. {\tt HIJING} always produces no correlation while {\tt MC-Glauber} produces correlations with a specific sign pattern.}
\label{fig:MHC_HIJING} 
\end{center}
\end{figure}

\begin{figure}[htbp!]
\begin{center}
\includegraphics[width=0.98\linewidth]{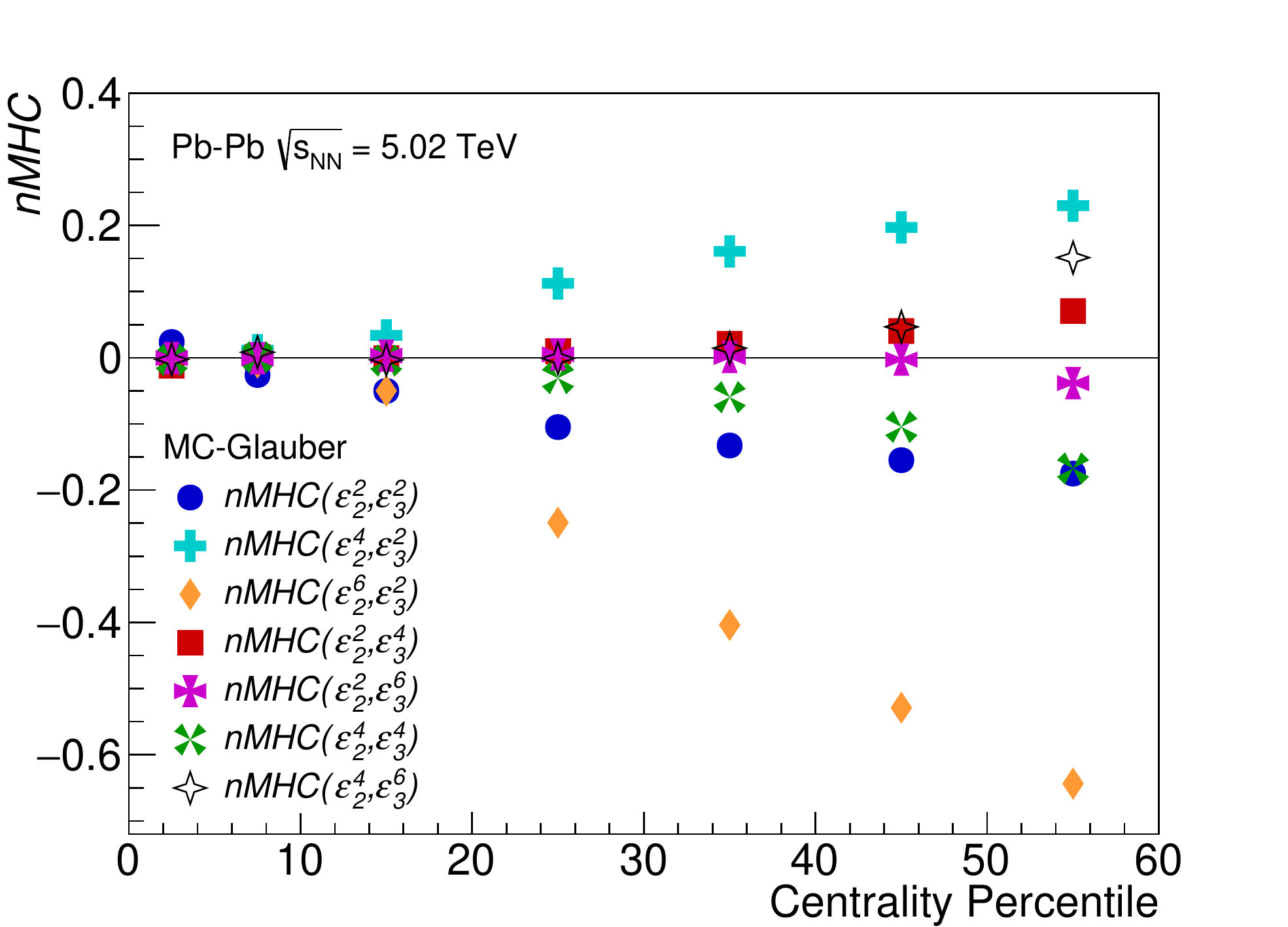}
\caption{(Color online) Centrality dependence of normalized mixed-harmonic cumulants ($nMHC$) in Pb--Pb collisions from the {\tt MC-Glauber} model. Stronger correlations are seen with increasing powers of $\varepsilon_2$.}
\label{fig3} 
\end{center}
\end{figure}

To study the effect of fluctuations we again employ the {\tt MC-Glauber} model. In order to provide reasonable predictions, we must scale $\varepsilon_2\{2\}$ and $\varepsilon_3\{2\}$ to $v_2\{2, |\Delta \eta| > 1\}$ and $v_3\{2, |\Delta \eta| > 1\}$ from ALICE~\cite{Adam:2016izf}. Then the calculations from the {\tt MC-Glauber} model can be directly compared to the measurements in the final state (i.e. the correlations between the flow coefficients).
The results are shown in Fig.~\ref{fig:MHC_HIJING}, where we see the characteristic negative, positive, and negative signs of 4-, 6-, and 8-particle mixed harmonic cumulants, respectively. This is, in fact, very similar to what has been observed in multi-particle cumulants of a single harmonic where positive, negative, positive, and negative signs of the 2-, 4-, 6-, and 8-particle cumulants of $v_2$~\cite{Acharya:2018lmh, Acharya:2019vdf} are observed. The different signs of mixed-harmonic cumulants further suggest that one should not conclude $v_2$ and $v_3$ are anti-correlated from the measured negative value of $SC(2,3)$. One should instead, more narrowly, only conclude that $v_2^2$ and $v_3^2$ are anti-correlated.  
%Meanwhile, if the {\tt MC-Glauber} model can describe the real correlations between different moments of $\varepsilon_2$ and $\varepsilon_3$ in the initial stage of heavy-ion collisions, then we would expect a good agreement between the prediction shown here and the experimental measurements to be available in the near future. 
Although it is possible that the {\tt MC-Glauber} model might not describe quantitatively the real correlations between different moments of $\varepsilon_2$ and $\varepsilon_3$ in the initial stage of heavy-ion collisions, we expect that the general trend of the centrality dependence as well as the hierarchy of different orders of $MHC$ should hold. Direct data and theory comparisons can be made via normalized mixed-harmonic cumulants, $nMHC$, because the scaling parameters $\kappa_2$ and $\kappa_3$ should cancel out. Thus the future experimental measurements of $nMHC$ could be used to directly constrain the initial state model in both large and small systems. The prediction of $nMHC$ using the {\tt MC-Glauber} model is shown in Fig.~\ref{fig3}. It is found that the correlations between higher moments of $v_2$ with $v_3$ are generally stronger than the correlations between low moments of $v_2$ and higher moments of $v_3$. To be more specific, $nMHC(\varepsilon_2^{6}, \varepsilon_3^2)$ has the largest anti-correlations, which is followed by $nMHC(\varepsilon_2^{4}, \varepsilon_3^2)$. It will be beneficial to check with actual experimental measurements whether the above ordering of the magnitudes of $nMHC$ persists.

 %All the above proposed $MHC$ are constructed based on multi-particle correlations, which can be obtained directly via the well established multi-particle generic framework~\cite{Bilandzic:2013kga,Huo:2017nms}. This is a framework that enables efficient and exact evaluations of multi-particle correlations in both theoretical calculations and experimental measurements, and has been widely used in almost all experiments at RHIC and the LHC. 
%Considering the statistics collected by the LHC experiments, we do not extend the list to higher order $MHC$ in this paper.

The correlations between flow harmonics, especially the symmetric cumulants, have provided important insight into collectivity in small collision systems. As a multi-particle cumulant, $MHC$ should suppress the non-flow contamination by construction. This insensitivity to non-flow has been tested in {\tt HIJING} simulations where all presented $MHC$ measurements are consistent with zero, as shown in Fig.~\ref{fig:MHC_HIJING}. Therefore, future experimental measurements of $MHC$ in small collision systems (where non-flow remains a significant contribution to flow measurements) could be valuable for constraining theoretical models, as currently both the Color Glass Condensate (CGC) and hydrodynamic models find it challenging to quantitatively or even qualitatively describe the experimental measurements of multi-particle cumulants. One recent example of this is the failure of hydrodynamic flow to describe the negative $c_2\{4\}$ and $SC(2,3)$ values in 13 TeV proton-proton collisions~\cite{Zhao:2020pty}. higher order mixed-harmonic cumulants will bring better sensitivity to constrain the models while reducing the non-flow bias in experimental measurements.

%not always in term of a strictly defined cumulant. It is mathematically incorrect for instance in the case of $v_2^2$, $v_3^2$ and $v_5^2$ correlations, where the non-vanishing term $\left<  v_2 \,v_3 \, v_5 \cos (2\Psi_2 + 3\Psi_3 - 5\Psi_5) \right>$ was mistakenly ignored.  As the so-called higher order symmetric cumulant is not the mathematically defined cumulant we define in this paper and less relevant to the discussions we have in the general formula for multi-particle cumulants we propose here, we will not comment further but leave our comments and discussions only in the appendix.

%We should add 10-pc with $MHC(v_2^8, v_3^2)$ (2,2,2,2,3,-2,-2,-2,-2,3), also $MHC(v_2^4, v_3^6)$ (2,2,3,3,3,-2,-2,-3,-3,-3) to confirm if they are positive again.

%For the {\tt MC-Glauber}  results, it is shown in Fig.~\ref{} that the negative, positive, negative and positive signs for multi-particle cumulants of MHC, this is originated from correlations with between various moments of anisotropy coefficients in the initial conditions. If the statistics allow, we might even be able to discuss the size of nMHC.

\section{Summary}
\label{sec:summary}

In this paper, a generic algorithm for multi-particle cumulants, for both single and mixed harmonics, has been proposed and studied using Monte Carlo simulations. The new 10-, 12-, 14-, and 16-particle cumulants of a single harmonic, named $c_{n}\{10\}$, $c_{n}\{12\}$, $c_{n}\{14\}$, and $c_{n}\{16\}$, as well as their corresponding $v_n$ coefficients, have been presented, which show different sensitivities to the event-by-event fluctuations compared to the 2-particle cumulant. %These new observables could significantly improve the probe of the initial conditions of the heavy-ion collisions. 
The higher order cumulants with more than 8-particles possess less non-flow contamination as shown in the MC study utilizing a {\tt HIJING} simulation. Therefore, future experimental measurements of these higher order cumulants could be beneficial in the understanding of the origins of flow in small collision systems which is currently under intense debate. In addition, the proposed mixed-harmonic cumulants ($MHC$) has been studied with both the {\tt HIJING} transport model and the {\tt MC-Glauber} model with proper scaling. It is found that negative, positive, and negative signs have been observed for 4-, 6-, and 8-particle mixed-harmonic cumulants. %We also observed that the correlations between $nMHC(v_2^k, v_3^l)$ show a characteristic pattern of negative, positive, negative signs from low to high order moment correlations, which agree with the corresponding calculations nicely in the initial condition simulated by {\tt MC-Glauber} model.
Compared to the standard individual $v_{n}$ coefficients of lower order cumulants and the previous symmetric cumulants, the high-order cumulants of a single harmonic and the mixed-harmonic cumulants proposed in this paper will provide new constraints for detailed tests of theoretical calculations. Future experimental measurements of the predicted observables will shed new light into the nature of initial state fluctuations and the properties of the QGP fireball created in the ultra-relativistic heavy-ion collisions. 

\section{Acknowledgment}\label{sec:Ack}

We thank Cvetan Cheshkov, Andrea Dainese, Alice Ohlson, Jean-Yves Ollitrault, Anthony Timmins, Sergei Voloshin and Flow PAG of ALICE Collaboration for very fruitful discussions. 
This work is supported by a research grant (00025462) from VILLUM FONDEN, the Danish National Research Foundation (Danmarks Grundforskningsfond), and the Carlsberg Foundation (Carlsbergfondet).

\appendix
\section{general formula for 6-particle cumulant}

For the future reference, we provide here the complete formula for a 6-particle cumulants without terms with only a single harmonic (as they must average to zero over many events).

\onecolumngrid

{ \scriptsize

\begin{eqnarray}
Cum(n_1, n_2, n_3, n_4, n_5, n_6) &=&  \left< \left< e^{ i \,(n_1\varphi_1 + n_2 \varphi_2 + n_3\varphi_3 + n_4\varphi_4  +n_5\varphi_5  + n_6\varphi_6  ) }\right>\right>_c  \nonumber\\
&=& \left< \left< e^{i \, (n_1\varphi_1 + n_2 \varphi_2 + n_3\varphi_3 + n_4\varphi_4  +n_5\varphi_5  + n_6\varphi_6  )}\right>\right>   \nonumber\\
&~& +  2*\left< \left< e^{ i \, (n_1\varphi_1+ n_6\varphi_6  )}\right>\right> \left< \left< e^{ i \, (n_4\varphi_4  +n_5\varphi_5 )}\right>\right> \left< \left< e^{i \, (n_2 \varphi_2 + n_3\varphi_3)} \right>\right>  \nonumber\\
&~& +  2*\left< \left< e^{i \, (n_1\varphi_1+ n_6\varphi_6  )}\right>\right> \left< \left<  e^{i \, (n_4\varphi_4  +n_2\varphi_2 )}\right>\right> \left< \left< e^{i \, (n_5 \varphi_5 + n_3\varphi_3)}\right>\right>  \nonumber\\
&~& +  2*\left< \left< e^{i \, (n_1\varphi_1+ n_6\varphi_6  )}\right>\right> \left< \left< e^{i \, (n_4\varphi_4  +n_3\varphi_3 )}\right>\right> \left< \left< e^{i \, (n_2 \varphi_2 + n_5\varphi_5)}\right>\right>  \nonumber\\
&~& +  2*\left< \left< e^{i \, (n_2\varphi_2+ n_6\varphi_6  )} \right>\right> \left< \left< e^{i \, (n_4\varphi_4  +n_1\varphi_1 )}\right>\right> \left< \left< e^{i \,(n_5 \varphi_5 + n_3\varphi_3)}\right>\right>  \nonumber\\
&~& +  2*\left< \left< e^{i \,(n_2\varphi_2+ n_6\varphi_6  )}\right>\right> \left< \left< e^{i \, (n_4\varphi_4  +n_5\varphi_5 )}\right>\right> \left< \left< e^{i \, (n_1 \varphi_1 + n_3\varphi_3)} \right>\right>  \nonumber\\
&~& +  2*\left< \left< e^{i \, (n_2\varphi_2+ n_6\varphi_6  )}\right>\right> \left< \left< e^{i \, (n_4\varphi_4  +n_3\varphi_3 )}\right>\right> \left< \left< e^{i \,(n_1 \varphi_1 + n_5\varphi_5)}\right>\right>  \nonumber\\
&~& +  2*\left< \left< e^{i \, (n_3\varphi_3+ n_6\varphi_6  )} \right>\right> \left< \left< e^{i \, (n_4\varphi_4  +n_1\varphi_1 )}\right>\right> \left< \left< e^{i \, (n_2 \varphi_2 + n_5\varphi_5) } \right>\right>  \nonumber\\
&~& +  2*\left< \left< e^{i \, (n_3\varphi_3+ n_6\varphi_6  )}\right>\right> \left< \left< e^{i \, (n_2\varphi_2  +n_4\varphi_4 )}\right>\right> \left< \left< e^{i \, (n_1 \varphi_1+ n_5\varphi_5)}\right>\right>  \nonumber\\
&~& +  2*\left< \left< e^{i \, (n_3\varphi_3+ n_6\varphi_6  )}\right>\right> \left< \left< e^{i \, (n_5\varphi_5  +n_4\varphi_4 )}\right>\right> \left< \left< e^{i \, (n_1 \varphi_1+ n_2\varphi_2)}\right>\right>  \nonumber\\
&~& +  2*\left< \left< e^{i \, (n_4\varphi_4+ n_6\varphi_6  ) } \right>\right> \left< \left< e^{i \, (n_5\varphi_5  +n_1\varphi_1 )}\right>\right> \left< \left< e^{i \, (n_2 \varphi_2 + n_3\varphi_3) } \right>\right>  \nonumber\\
&~& +  2*\left< \left< e^{i \, (n_4\varphi_4+ n_6\varphi_6  )} \right>\right> \left< \left< e^{i \, (n_5\varphi_5  +n_2\varphi_2 )}\right>\right> \left< \left< e^{i \, (n_1 \varphi_1 + n_3\varphi_3)}\right>\right>  \nonumber\\
&~& +  2*\left< \left< e^{i \, (n_4\varphi_4+ n_6\varphi_6  ) } \right>\right> \left< \left< e^{i \,  n_5\varphi_5  +n_3\varphi_3 )}\right>\right> \left< \left< e^{i \, (n_1 \varphi_1 + n_2\varphi_2)} \right>\right>  \nonumber\\
&~& +  2*\left< \left< e^{i \, (n_5\varphi_5+ n_6\varphi_6  )} \right>\right> \left< \left< e^{i \, n_2\varphi_2  +n_3\varphi_3 )}\right>\right> \left< \left< e^{i \, (n_1 \varphi_1 + n_4\varphi_4)}\right>\right>  \nonumber\\
&~& +  2*\left< \left< e^{i \,(n_5\varphi_5+ n_6\varphi_6  ) }\right>\right> \left< \left< e^{ i \, ( n_2\varphi_2  +n_4\varphi_4 ) }\right>\right> \left< \left< e^{i \, (n_1 \varphi_1 + n_3\varphi_3) }\right>\right>  \nonumber\\
&~& +  2*\left< \left< e^{i \, (n_5\varphi_5+ n_6\varphi_6  ) }\right>\right> \left< \left< e^{ i \, ( n_4\varphi_4  +n_3\varphi_3 ) }\right>\right> \left< \left< e^{i \, (n_1 \varphi_1 + n_2\varphi_2) }\right>\right>  \nonumber\\
&~& - 1*\left< \left< e^{i \, (n_1\varphi_1+ n_6\varphi_6  ) }\right>\right> \left< \left< e^{ i \, ( n_2\varphi_2  +n_3\varphi_3  + n_4 \varphi_4 + n_5\varphi_5) }\right>\right>  \nonumber\\
&~& - 1*\left< \left< e^{i \, (n_2\varphi_2+ n_6\varphi_6  ) }\right>\right> \left< \left< e^{ i \, ( n_1\varphi_1  +n_3\varphi_3  + n_4 \varphi_4 + n_5\varphi_5) }\right>\right>  \nonumber\\
&~& - 1*\left< \left< e^{i \, (n_3\varphi_3+ n_6\varphi_6  ) }\right>\right> \left< \left< e^{ i \, ( n_1\varphi_1  +n_2\varphi_2  + n_4 \varphi_4 + n_5\varphi_5) }\right>\right>  \nonumber\\
&~& - 1*\left< \left< e^{i \, (n_4\varphi_4+ n_6\varphi_6  ) }\right>\right> \left< \left< e^{ i \, ( n_1\varphi_1  +n_2\varphi_2  + n_3 \varphi_3 + n_5\varphi_5) }\right>\right>  \nonumber\\
&~& -1 *\left< \left< e^{i \, (n_5\varphi_5+ n_6\varphi_6  ) }\right>\right> \left< \left< e^{ i \, ( n_1\varphi_1  +n_2\varphi_2  + n_3 \varphi_3 + n_4\varphi_4) }\right>\right>  \nonumber\\
&~& - 1*\left< \left< e^{i \, (n_1\varphi_1 + n_2\varphi_2 + n_6\varphi_6  ) }\right>\right> \left< \left< e^{ i \, ( n_3 \varphi_3 + n_4\varphi_4 + n_5\varphi_5 ) }\right>\right>  \nonumber\\
&~& - 1*\left< \left< e^{i \, (n_1\varphi_1 + n_3\varphi_3 + n_6\varphi_6  ) }\right>\right> \left< \left< e^{ i \, ( n_2 \varphi_2 + n_4\varphi_4 + n_5\varphi_5 ) }\right>\right>  \nonumber\\
&~& - 1*\left< \left< e^{i \, (n_1\varphi_1 + n_4\varphi_4 + n_6\varphi_6  ) }\right>\right> \left< \left< e^{ i \, ( n_3 \varphi_3 + n_2\varphi_2 + n_5\varphi_5 ) }\right>\right>  \nonumber\\
&~& - 1*\left< \left< e^{i \, (n_1\varphi_1 + n_5\varphi_5 + n_6\varphi_6  ) }\right>\right> \left< \left< e^{ i \, ( n_3 \varphi_3 + n_2\varphi_2 + n_4\varphi_4 ) }\right>\right>  \nonumber\\
&~& -1 *\left< \left< e^{i \, (n_2\varphi_2 + n_3\varphi_3 + n_6\varphi_6  ) }\right>\right> \left< \left< e^{ i \, ( n_1 \varphi_1 + n_4\varphi_4 + n_5\varphi_5 ) }\right>\right>  \nonumber\\
&~& - 1*\left< \left< e^{i \, (n_2\varphi_2 + n_4\varphi_4 + n_6\varphi_6  ) }\right>\right> \left< \left< e^{ i \, ( n_1 \varphi_1 + n_3\varphi_3 + n_5\varphi_5 ) }\right>\right>  \nonumber\\
&~& - 1*\left< \left< e^{i \, (n_2\varphi_2 + n_5\varphi_5 + n_6\varphi_6  ) }\right>\right> \left< \left< e^{ i \, ( n_1 \varphi_1 + n_3\varphi_3 + n_4 \varphi_4 ) }\right>\right>  \nonumber\\
&~& - 1*\left< \left< e^{i \, (n_3\varphi_3 + n_4\varphi_4 + n_6\varphi_6  ) }\right>\right> \left< \left< e^{ i \, ( n_1 \varphi_1 + n_2\varphi_2 + n_5\varphi_5 ) }\right>\right>  \nonumber\\
&~& - 1*\left< \left< e^{i \, (n_3\varphi_3 + n_5\varphi_5 + n_6\varphi_6  ) }\right>\right> \left< \left< e^{ i \, ( n_1 \varphi_1 + n_2\varphi_2 + n_4\varphi_4 ) }\right>\right>  \nonumber\\
&~& - 1*\left< \left< e^{i \, (n_4\varphi_4 + n_5\varphi_5 + n_6\varphi_6  ) }\right>\right> \left< \left< e^{ i \, ( n_1 \varphi_1 + n_2\varphi_2 + n_3\varphi_3 ) }\right>\right>  \nonumber\\
&~& - 1*\left< \left< e^{i \, (n_1\varphi_1+ n_2\varphi_2  ) }\right>\right> \left< \left< e^{ i \, ( n_3\varphi_3  +n_4\varphi_4 + n_5\varphi_5 + n_6\varphi_6) }\right>\right>  \nonumber\\
&~& - 1*\left< \left< e^{i \, (n_1\varphi_1+ n_3\varphi_3  ) }\right>\right> \left< \left< e^{ i \, ( n_2\varphi_2  +n_4\varphi_4 + n_5\varphi_5 + n_6\varphi_6) }\right>\right>  \nonumber\\
&~& -1 *\left< \left< e^{i \, (n_1\varphi_1+ n_4\varphi_4  ) }\right>\right> \left< \left< e^{ i \, ( n_3\varphi_3  +n_2\varphi_2 + n_5\varphi_5 + n_6\varphi_6) }\right>\right>  \nonumber\\
&~& -1 *\left< \left< e^{i \, (n_1\varphi_1+ n_5\varphi_5  ) }\right>\right> \left< \left< e^{ i \, ( n_3\varphi_3  +n_2\varphi_2 + n_4\varphi_4 + n_6\varphi_6) }\right>\right>  \nonumber\\
&~& -1 *\left< \left< e^{i \, (n_2\varphi_2+ n_3\varphi_3  ) }\right>\right> \left< \left< e^{ i \, ( n_1\varphi_1 +n_4\varphi_4 + n_5\varphi_5 + n_6\varphi_6) }\right>\right>  \nonumber\\
&~& - 1*\left< \left< e^{i \, (n_2\varphi_2+ n_4\varphi_4  ) }\right>\right> \left< \left< e^{ i \, ( n_1\varphi_1 +n_3\varphi_3 + n_5\varphi_5 + n_6\varphi_6) }\right>\right>  \nonumber\\
&~& -1 *\left< \left< e^{i \, (n_2\varphi_2+ n_5\varphi_5  ) }\right>\right> \left< \left< e^{ i \, ( n_1\varphi_1 +n_3\varphi_3 + n_4\varphi_4 + n_6\varphi_6) }\right>\right>  \nonumber\\
&~& -1 *\left< \left< e^{i \, (n_3\varphi_3+ n_4\varphi_4  ) }\right>\right> \left< \left< e^{ i \, ( n_1\varphi_1 +n_2\varphi_2 + n_5\varphi_5 + n_6\varphi_6) }\right>\right>  \nonumber\\
&~& - 1*\left< \left< e^{i \, (n_3\varphi_3+ n_5\varphi_5  ) }\right>\right> \left< \left< e^{ i \, ( n_1\varphi_1 +n_2\varphi_2 + n_4\varphi_4 + n_6\varphi_6) }\right>\right>  \nonumber\\
&~& - 1*\left< \left< e^{i \, (n_4\varphi_4+ n_5\varphi_5  ) }\right>\right> \left< \left< e^{ i \, ( n_1\varphi_1 +n_2\varphi_2 + n_3\varphi_3 + n_6\varphi_6) }\right>\right>  
\label{eq:6pc_6PC}
\end{eqnarray}

}

 \twocolumngrid

\section{higher order symmetric cumulant}
\label{appb}

It should be pointed out that, in some cases, multi-particle mixed-harmonic cumulants contain non-vanishing terms with event-plane correlations. For example,

{\small

\begin{eqnarray}
&& ~~ \left< \left<  e^{ i ( 2\varphi_1 + 3\varphi_2 + 5 \varphi_3 - 2\varphi_4 - 3\varphi_5 - 5\varphi_6)} \right> \right>_c = \nonumber\\
&& \left< \left< e^{ i (2\varphi_1 + 3\varphi_2 + 5 \varphi_3 - 2\varphi_4 - 3\varphi_5 - 5\varphi_6)} \right> \right>   \nonumber\\
& & -  \left< \left< e^{ i (2\varphi_1 + 3\varphi_2 - 2\varphi_3 - 3\varphi_4)} \right> \right> \, \left< \left< e^{ i  ( 5 \varphi_1 - 5\varphi_2)} \right> \right> \nonumber\\
& &-  \left< \left< e^{ i  (2\varphi_1 + 5\varphi_2 - 2\varphi_3 - 5\varphi_4) } \right> \right> \, \left< \left< e^{ i  ( 3 \varphi_1 - 3\varphi_2)} \right> \right> \nonumber\\
& & -  \left< \left< e^{ i  (3\varphi_1 + 5\varphi_2 - 3\varphi_3 - 5\varphi_4)} \right> \right> \, \left< \left< e^{ i  ( 2 \varphi_1 - 2\varphi_2)} \right> \right> \nonumber\\
& & \bf{-  \left< \left< e^{ i  (2\varphi_1 + 3\varphi_2 - 5\varphi_3) }\right> \right> \, \left< \left< e^{ i ( 2 \varphi_1 + 3 \varphi_2 - 5\varphi_3)} \right> \right>} \nonumber\\
& &  +  2 \left< \left< e^{ i  (2\varphi_1 - 2\varphi_2 )} \right> \right>  \left< \left< e^{ i (3\varphi_1 - 3\varphi_2 )} \right> \right>  \left< \left< e^{ i (5\varphi_1 - 5\varphi_2 )} \right> \right>  
\end{eqnarray} 
}

It is clear that the non-vanishing term $ \left< \left< e^{(2\varphi_1 + 3\varphi_2 - 5\varphi_3)} \right> \right>$ has the event-plane correlation between $\Psi_2$, $\Psi_3$ and $\Psi_5$.

It is noticed that in Ref.~\cite{Mordasini:2019hut}, the authors proposed a similar idea as $MHC$ to study the correlations between multiple flow coefficients. However, the proposed generation of higher order cumulants for 6-particles is based on Kubo's three observable cumulants, which treats $v_n^2$ as a fundamental observable. Considering the existing flow technique is still based on the azimuthal correlation, it is not obvious how to obtain $v_n^2$ event-by-event in experimental measurements. In addition, the lower-order azimuthal correlations might not be removed completely (e.g., in case of $n_1 = 2, n_2 = 3, n_3 = 5$ discussed above), the remaining non-flow effects could be more significant than the multi-particle cumulants defined using azimuthal angle as the fundamental observable. Last but not least, the proposal of `shift of paradigm' using $v_n^2$ as an observable will lead to the fact that one will reconstruct a simple four-particle cumulant of single harmonic with: $c_{n}\{4\} = \left< v_n^4 \right> - \left< v_n^2 \right>^2$, instead of standard $c_{n}\{4\} = \left< v_n^4 \right> - 2 \, \left< v_n^2 \right>^2$. The latter clearly suppresses non-flow better, which has been studied for two decades~\cite{Borghini:2000sa}. In general, it is the same case for any 6-particle cumulant $\left< \left< \cos (k\varphi_1 + l\varphi_2 + m \varphi_3 - k\varphi_4 - l\varphi_5 - m\varphi_6) \right> \right>_c$ when $k+l-m =0$. 

%While we have less doubt that "shift of paradigm" might be potentially interesting, we would still emphasis that it is extremely challenging to obtained $v_n^2$ event-by-event in experiments, using currently existing technique that starts from azimuthal angle of emitted particle. And it is also obvious that for some observables, the non-flow contaminations might be non-negligible as the lower order azimuthal correlations might not be removed completely. We will not comment further on 8-particle cumulants as it will have many more non-vanishing event-plane correlations, which are not included in the current implementation of higher order symmetric cumulant. 

\bibliography{bibliography}{}
\bibliographystyle{apsrev4-1}

\end{document}